\documentclass[12pt]{article}
\usepackage{a4wide}
\usepackage{amssymb, amsmath,cite}
\usepackage[normalem]{ulem}
\usepackage[utf8]{inputenc}
\usepackage[colorlinks]{hyperref}
\usepackage[usenames,dvipsnames]{color}
\hypersetup{
     breaklinks=true,
    pdfstartview={FitH},    % fits the width of the page to the window
    colorlinks=true,       % false: boxed links; true: colored links
    linkcolor=blue,          % color of internal links
    citecolor=red,        % color of links to bibliography
    filecolor=magenta,      % color of file links
    urlcolor=blue,           % color of external links
    anchorcolor=green,      % Color for anchor text
    linktocpage=true
}

\begin{document}
{\renewcommand{\thefootnote}{\fnsymbol{footnote}}
		
\begin{center}
{\LARGE Trans-Planckian censorship, inflation and excited initial states for perturbations} 
\vspace{1.5em}

Suddhasattwa Brahma\footnote{e-mail address: {\tt suddhasattwa.brahma@gmail.com}}
\\
\vspace{0.5em}
 Department of Physics, McGill University\\
 Montr\'eal, QC H3A 2T8, Canada\\
\vspace{1.5em}
\end{center}
}
	
\setcounter{footnote}{0}

\newcommand{\bea}{\begin{eqnarray}}
\newcommand{\eea}{\end{eqnarray}}
\renewcommand{\d}{{\mathrm{d}}}
\renewcommand{\[}{\left[}
\renewcommand{\]}{\right]}
\renewcommand{\(}{\left(}
\renewcommand{\)}{\right)}
\newcommand{\nn}{\nonumber}
\newcommand{\Mpl}{M_{\textrm{Pl}}}
\def\H{\mathrm{H}}
\def\V{\mathrm{V}}
\def\e{\mathrm{e}}
\def\be{\begin{equation}}
\def\ee{\end{equation}}

\begin{abstract}
\noindent The recently proposed \textit{trans-Planckian censorship conjecture}  (TCC) seems to require that the energy scale of inflation is significantly lower than the Planck scale $(H_\text{inf}<10^{-20} \Mpl)$. This, in turn, implies that the tensor-to-scalar ratio for inflation is negligibly small, \textit{independent} of assumptions of slow-roll or even of having a single scalar field, thus ruling out inflation if primordial tensor modes are ever observed. After demonstrating the robustness and generality of these bounds, we show that having an excited initial state for cosmological perturbations seems to be a way out of this problem for models of inflation.
\end{abstract}

\section{TCC \& Inflation}
One of the crowning glories of inflation \cite{Inf1, Inf2, Inf3, Inf4, Inf5,Inf6} lies in its ability to explain the origin of small inhomogeneities, which seed the large scale structure of our universe, as quantum vacuum fluctuations \cite{Pert1, Pert2}. As is well-known \cite{Review1, Review2}, these quantum fluctuations which oscillate within the Hubble horizon, become classical and freeze on exiting it during inflation, and eventually re-enter the horizon at late times to seed the anisotropies in the cosmic microwave background radiation \cite{WMAP, Planck1} and the large scale structure \cite{SDSS} which we observe  today. 

However, using the above mechanism, one can easily see that even trans-Planckian quantum fluctuations can become classical if inflation lasts long enough (and, in particular, much longer than what is needed to explain our current observations) \cite{TP1}. In other words, inflation, as a low-energy effective field theory (EFT), would fail if macroscopic perturbations can be traced backed to modes which are smaller than the Planck length at early times. This is the well-known \textit{`trans-Planckian problem'} in cosmology \cite{TP1,TP2} (see also, \cite{TP3,TP4,TP5,TP6,TP7,TP8}) and it was a general expectation that the details of the theory behind the quantum completion of inflation would resolve it. 

Instead, recently it has been conjectured \cite{TCC1} that in any consistent theory of gravity, quantum fluctuations smaller than the Planck length should never be able to exit the Hubble horizon and become classical. This \textit{`trans-Planckian censorship conjecture'} (TCC) adds to the list of the consistency requirements  -- the so-called swampland conjectures \cite{Swampland1,Swampland2,Swampland3} -- for a quantum theory of gravity. The TCC seems similar in spirit to Penrose's `Cosmic Censorship hypothesis' \cite{Penrose1,Penrose2}, where in place of a singularity, fluctuations in the trans-Planckian regime are conjectured to be hidden by the Hubble horizon\footnote{The author is thankful to Robert Brandenberger for this point of view.}.

As has been pointed out in \cite{TCC1,TCC2}, the TCC can only be violated for accelerated expansion and is, therefore, in some tension with regimes such as inflation (and does not pose a problem for standard radiation or matter-dominated eras). In \cite{TCC2}, the implications of the TCC for inflationary cosmology has been studied in some detail (see \cite{TCC_Cosmo1,TCC_Cosmo2,TCC_Cosmo3,TCC_Cosmo4, TCC_Cosmo5,TCC_Cosmo6} for related work). Let us first summarize their main results and demonstrate the robustness of their findings.

As mentioned earlier, quantum mechanical fluctuations oscillate on sub-horizon scales but freeze out when their wavelengths become larger than the Hubble radius \cite{Review2}. Therefore, during accelerated expansion, sub-horizon modes can exit the horizon and become classical, presumably through some decoherence mechanism \cite{Q-to-C1,Q-to-C2,Q-to-C3,Q-to-C4,Q-to-C5}. Since the TCC prohibits this for trans-Planckian modes, one gets an upper bound on the duration of such accelerated phases so that a Planck length mode never exits the Hubble horizon. For inflation, this implies \cite{TCC2}
\begin{eqnarray}
e^N < \Mpl/H_\text{inf}\,,
\end{eqnarray}
where $N$ is the number of $e$-folds of inflation and we assume that the Hubble parameter $H_\text{inf}$ remains constant during inflation. On the other hand, in order to retain its most striking success, inflation needs to explain the scales we see today as originating from quantum fluctuations, \textit{i.e.} the comoving Hubble radius today must have been sub-horizon at the beginning of inflation to be in causal contact. This puts a lower bound on inflation \cite{TCC2}.  These two bounds sets up an hierarchy for $N$ and, requiring that they are compatible, results in an upper bound on the energy density for inflation \cite{TCC2}
\begin{eqnarray}\label{Eq1}
\rho_\text{inf}^{1/4} < 3 \times 10^{-10} \Mpl\,.
\end{eqnarray}
This conclusion is completely independent of the details of inflation and sets an upper bound on the energy density of the background during inflation from the Friedmann equation alone\footnote{There have been some simplifying assumptions in this order of magnitude derivation (see \cite{TCC2} for details) but none so severe to change this result drastically.}.  

Next the authors of \cite{TCC2} puts an upper bound on the first slow-roll parameter $\epsilon$, defined as $3/2\left(1+ p/\rho\right)$ by using the Hubble scale during inflation and the observed value of the curvature power spectrum, from the equation
\begin{eqnarray}\label{Eq2}
\mathcal{P}_\mathcal{R} = \frac{1}{8\pi^2\epsilon} \left(\frac{H_\text{inf}}{\Mpl}\right)^2\,,
\end{eqnarray}
by approximating $\mathcal{P}_\mathcal{R} \sim 10^{-9}$, and using the maximum allowed value for $H_\text{inf} \sim 3^{1.5} \times 10^{-20} \Mpl$ from \eqref{Eq1}. This gives the bound 
\begin{eqnarray}\label{Eq3}
\epsilon < 10^{-31} \,.
\end{eqnarray}
Finally, on using the consistency relation $r = 16\epsilon$, where $r$ is the tensor-to-scalar ratio, one gets the final bound in \cite{TCC2}
\begin{eqnarray}\label{r-Eqn}
r < 10^{-30}\,.
\end{eqnarray}
This is the most remarkable of all the constraints since it implies that the detection of primordial gravitational waves would \textit{rule out} inflation, assuming the TCC. 

As was already noted in \cite{TCC2}, we stress that these bounds do not assume any slow-roll condition for their derivation. In fact, one can clearly see the robustness in this derivation of the bound on $r$ in the following way. At first, it might seem that since the bound on $r$ follows from the bound on $\epsilon$ \eqref{Eq3}, it might be easy to avoid it by postulating some model of inflation which violates the $r = 16\epsilon$ consistency relation. However, that is not the case since the constraint \eqref{Eq3} is not so much on $\epsilon$, but rather on the factor multiplying $H_\text{inf}/M_{pl}$ in the scalar power spectrum. Any mechanism, such as going to multi-fields or warm inflation, which modifies this consistency relation by modifying the scalar power spectrum would not be able to affect the bound on $r$ \eqref{r-Eqn}. To explicitly demonstrate this, let us consider some incarnation of inflation which leads to the following expression for the curvature power spectrum
\begin{eqnarray}\label{Robust1}
\mathcal{P}_\mathcal{R} = \frac{1}{8\pi^2\epsilon} \left(\frac{H_\text{inf}}{\Mpl}\right)^2 \Gamma_s\,,
\end{eqnarray}
where $\Gamma_s$ is the model-dependent modification factor. This would lead to a new consistency relation 
\begin{eqnarray}\label{Robust2}
r = 16 \epsilon/\Gamma_s\,.
\end{eqnarray}
However, plugging in $\mathcal{P}_\mathcal{R} \sim 10^{-9}$ and $H_\text{inf} < 3^{1.5} \times 10^{-20} \Mpl$ from \eqref{Eq1} gives a constraint on $\epsilon/\Gamma_s < 10^{-31}$. Since this is the precise combination which appears in \eqref{Robust2}, it means that the bound on $r<10^{-30}$ would remain unaffected. Indeed, for the case of warm inflation, this has been recently verified in \cite{TCC_Cosmo3} (\textit{Per se}, none of the models of inflation are in conflict with the TCC, but the real danger is that they would get ruled out if primordial tensor modes are ever observed. This last conclusion is also true for warm inflation and, thus, as far as the TCC is concerned, warm inflation models fare no better than their cold counterparts \cite{TCC2}.) Thus, from this simple calculation, it immediately follows that even if the bound on $\epsilon$ can be somewhat loosened in more sophisticated models of inflation going beyond single-field vanilla slow-roll, the bound on $r$ is more robust and is more generally applicable for all models of inflation. 

This, of course, also gives us pointers for what is needed for evading this bound. The first obvious option would be to give up on inflation and consider an alternative early-universe scenario (see \cite{Alternative} for a review) which is naturally compatible with the TCC, \textit{i.e.} it does not require any fine-tuning to satisfy the TCC, and is consistent with current observations. In other words, assuming that we detect primordial gravitational waves in the future, it might be time to consider a mechanism other than inflation to explain its origins, having assumed the TCC to be true. However, if we want to salvage inflation from this conundrum, one option would be to posit some departure from standard Big-Bang cosmology \textit{after} the inflationary era\footnote{As we were in the final stages of preparing this draft, \cite{TCC_Cosmo2} appeared on arXiv which seems to have gone this way in loosening the bounds of \cite{TCC2}.}. For instance, if one demands that there are large phases of expansion for which the equation of state \textit{does not} obey $w \geq 1/3$, then it might be able to weaken the bound \eqref{Eq1}. Naturally, if it was possible to somehow increase the upper bound on the energy density of inflation \eqref{Eq1}, the bound on $r$ \eqref{r-Eqn}, can also be consequently evaded. However, requiring such an \textit{ad hoc} departure from Big-Bang cosmology seems to be a rather radical assumption and shall not be pursued by us.

Another avenue for evading the bound \eqref{r-Eqn} would be to posit a new mechanism for production of tensor modes than what is considered in inflation. However, we want to do this in a way so as to not ruin the beauty of the theory as sourcing macroscopic inhomogeneities by quantum fluctuations, \textit{but not necessarily taken in the Bunch-Davies (BD) vacuum}. Pursuing this route, we show that there does remain one possibility of evading the bound by considering an excited initial state for cosmological perturbations with a non-Bunch Davies (NBD) component. Before going into the details of how this can help us in evading the bound, let us first provide some justifications for it.

Typically, it is assumed that both the scalar and the tensor modes are in the BD vacuum \cite{BD1,BD2} at the onset of inflation. One way to think of the BD state is to consider a mode at present time and then \textit{blueshift} it back all the way to the infinite past where it is far inside the horizon, such that it `feels' as if it is in flat space, and thus we can choose the ``Minkowski'' vacuum for it. On applying this procedure to all the modes, we arrive at the BD vacuum for de-Sitter (dS) space. However, if there is a cut-off in the theory, say the Planck scale $\Mpl$, then it would make little sense to blueshift a mode beyond this cut-off. Indeed, this is just a restatement of the `trans-Planckian' problem mentioned earlier \cite{TP8,AlphaVacuum}. In that case, the natural option would be to choose an \textit{instantaneous Minkowski vacuum}, as advocated in \cite{TP8,QuantumSwampland}. This leads to each of the modes being a Bogoliubov rotation of the BD mode, with a mixing between the creation and annihilation operators. More generally, if inflation started off at some finite time and is not past infinite, the pre-inflationary dynamics naturally points towards a NBD state for the fluctuations\footnote{In this paper, by NBD we shall always mean Bogoliubov transformation of the BD state and will not consider more general initial states such as mixed states or even non-Gaussian ones \cite{Agarwal:2012mq}.}, \textit{e.g.}  from a previous radiation-dominated era \cite{Vilenkin:1982wt}, a phase of anisotropic expansion \cite{Dey:2011mj,Dey:2012qp}, a non-attractor solution \cite{Lello:2013mfa}, tunneling from a false vacuum \cite{Sugimura:2013cra}, the effects of having a  high-energy cut-off \cite{TP1, Kempf:2000ac,Kaloper:2002uj, Hui:2001ce, Schalm:2004qk, Ashoorioon:2004wd}, multi-field dynamics \cite{Shiu:2011qw} or a specific quantum gravity proposal \cite{Agullo:2015aba}.  

In a nutshell, the TCC clearly prohibits us in assuming that the EFT description of a scalar field on an FLRW background is valid beyond the Planck scale and necessitates us to think about a pre-inflationary phase with inflation starting at a finite time. Therefore, it does not seem reasonable to assume the BD state at the onset of inflation, whose definition requires us to go to arbitrarily short distances \cite{TP8,Collins:2005nu}. This, in turn, naturally leads us towards a NBD state for inflation and we now show how assuming such a state might be able to evade the constraints given in \cite{TCC2}.

\section{Non-Bunch Davies initial states}
As mentioned, our ignorance regarding the very early-time physics shall be parametrized by the NBD intial states, for both the scalar ($\zeta$) and tensor ($h$) perturbations. Closely following the conventions of \cite{Brahma:2013rua}, we can write them as
\begin{eqnarray}
\zeta_{\bf k}(\eta) &=& v_k^{(s)} (\eta) a_{\bf k} + v_k^{(s)\star} (\eta) a_{\bf k}^\dagger,\\
h_{\bf k}^p(\eta) &=& v_k^{(t)} (\eta) a_{\bf k}^p + v_k^{(t)\star} (\eta) a_{\bf k}^{p\dagger},
\end{eqnarray}
where $a_{\bf k}, a_{\bf k}^\dagger$ stand for the usual creation and annihilation operators and $p$ labels graviton polarization. The $v_k^{(s,t)}$ and $u_k^{(s,t)}$\footnote{$(s)$ and $(t)$ stand for the scalar and tensor modes respectively} stand for the NBD and the BD modes respectively, and are related by
\begin{eqnarray}
v^{(s,t)}_k(\eta) = \alpha_k^{(s,t)} u_k^{(s,t)}(\eta) + \beta_k^{(s,t)}u^{(s,t)}_k(\eta)\,,
\end{eqnarray}
with the normalization condition for the Bogoliubov coefficients given by
\begin{eqnarray}
|\alpha_k^{(s,t)}|^2 - |\beta_k^{(s,t)}|^2 = 1\,.
\end{eqnarray}
Defined in terms of the two-point function in momentum space, the expression for the scalar and tensor power spectra is given by
\begin{eqnarray}
\langle \zeta_{\bf k_1} \zeta_{\bf k_1} \rangle &=& (2\pi) \delta^3({\bf k_1}+{\bf k_2}) P_\zeta(k_1)\,,\\
\langle h^p_{\bf k_1} h^{p^\prime}_{\bf k_1} \rangle &=& (2\pi) \delta^3({\bf k_1}+{\bf k_2}) \delta^{pp'} P^p_h(k_1)\,.
\end{eqnarray}
Writing out the explicit expressions for the  BD scalar and tensor modes,
\begin{eqnarray}
u^{(s)}_k(\eta) &=& \frac{H^2}{\dot{\varphi}} \frac{1}{\sqrt{2 k^3}} (1+ik\eta) e^{-ik\eta},\\
u^{(t)}_k(\eta) &=& \frac{H^2}{M_p} \frac{1}{\sqrt{ k^3}} (1+ik\eta) e^{-ik\eta}\,,
\end{eqnarray}
one readily gets the (dimensionless) power spectra as 
\begin{eqnarray}
\mathcal{P}_\mathcal{R} = \frac{1}{8\pi^2\epsilon} \left(\frac{H_\text{inf}}{\Mpl}\right)^2 \gamma_s\,,\label{NBD_PS_Scal} \\
\mathcal{P}_h = \frac{2}{\pi^2}\left(\frac{H_\text{inf}}{\Mpl}\right)^2 \gamma_t\,,\label{NBD_PS_Tensor}
\end{eqnarray}
where we have switched notation from $\zeta$ to $\mathcal{R}$ to be consistent with \eqref{Eq2} (and \cite{TCC2}). The modification factors are given by \cite{Brahma:2013rua,Ganc:2011dy}
\begin{eqnarray}
\gamma_{(s,t)} := |\alpha^{(s,t)}_k + \beta_k^{(s,t)}|^2 =  1 + 2 N_k^{(s,t)} + 2 \sqrt{N^{(s,t)}_k \left(1+N^{(s,t)}_k\right)} \cos\Theta_k^{(s,t)}\,,
\end{eqnarray}
with $N_k^{(s,t)}$ being the number of particles in a mode $k$ and $\Theta_k^{(s,t)}$, being the relative phase factor between the coefficients $\alpha_k^{(s,t)}$ and $\beta_k^{(s,t)}$, determines whether there is an enhancement or suppression to the BD value due to the excited states. 

Given these expressions, it is easy to calculate the tensor-to-scalar ratio as 
\begin{eqnarray}\label{r_NBD}
r = 16 \epsilon \frac{\gamma_{(t)}}{\gamma_{(s)}}\,.
\end{eqnarray}
Before going on to the phenomenology of these extra factors coming from NBD states \cite{Agarwal:2012mq,Hui:2001ce,Brahma:2013rua,Ganc:2011dy,Holman:2007na,Flauger:2013hra,Ashoorioon:2010xg,Ashoorioon:2013eia,Kundu:2011sg,Agullo:2010ws,Aravind:2014axa,Aravind:2013lra,Ashoorioon:2018sqb,Brahma:2018hrd,Shankaranarayanan:2002ax,Shankaranarayanan:2004iq,Sriramkumar:2006qt}, let us first sketch the argument behind why our mechanism might be able to evade the bound \eqref{r-Eqn}. Applying the bound \eqref{Eq1} and using the observed value of the power spectrum, we shall get a new bound, analogous to \eqref{Eq3}, given by
\begin{eqnarray}\label{Epsilon_NBD}
\epsilon/\gamma_{(s)} <10^{-31}\,.
\end{eqnarray}
However, in our case, since we assume that the initial state of both tensor and scalar perturbations would be modified in an \textit{independent} way, $r$ receives an extra modification from the tensor power spectrum. This way the bound on $r$ can be significantly reduced by utilizing the parameter space of NBD states involving excited tensor modes. At this point, let us emphasize the main assumption of our work: there are three different `clocks' for our model \cite{Brahma:2013rua} -- one for the background, one for the scalar perturbations and one for the tensors. Less effort has been put into concrete examples of generating general initial states for tensors as opposed to scalar perturbations. To the best of our knowledge, the only known physical mechanisms which produces a NBD state for tensors appears in theories where gravity is modified in the UV \cite{Horava:2009uw, Agullo:2015aba}, sometimes via modified dispersion relations \cite{Ashoorioon:2017toq}, even if one expects them to be present on more general considerations of EFT of fluctuations \cite{Holman:2007na,Agarwal:2012mq}. Another way to realize such NBD tensor modes might be to consider a thermal state for the fluctuations \cite{Bhattacharya:2005wn,Gasperini:1993yf} and this would be explored in the future. 

Having set up our general goal, let us see if we can achieve the required fine-tuning to sufficiently enhance the upper bound on $r$ coming from NBD states. The easiest approach would be to set the scalar modes in their BD vacuum ($\gamma_{(s)} = 1$) and only consider excited states for the tensor modes. However, the parameter $\gamma_{(t)}$ is, of course, not a completely free one and must respect backreaction and non-Gaussianity bounds. Following \cite{Holman:2007na}, we choose a model for the excited tensor and scalar modes as\footnote{Although we assume that the scalar modes are in the BD vacuum at this point, we anticipate the need for NBD scalar modes later on and include their definition at this point.}
\begin{eqnarray}\label{beta}
\beta_k^{(s,t)} \sim \beta_0^{(s,t)} e^{-k^2/\(M_{(s,t)} a(\eta_0)\)^2}\,,
\end{eqnarray}
where one assumes that at $\eta_0$, all of the observed primordial modes are below the cut-off scale $M_{(s,t)}$, \textit{i.e.} $k/a(\eta_0) \leq M_{(s,t)}$. We can choose the cut-off scale for the scalar and tensor modes to be the same, but we keep things general for now. For the usual EFT of slow-roll inflation to be valid, we must require that $M_{(s,t)} > H_\text{inf}$ and that $\beta^{(s,t)}(k) \rightarrow 0$ fast enough. This latter condition is the backreaction constraint which, for our parametrization \eqref{beta}, can be stated as \cite{Holman:2007na}\footnote{This is the stronger of the two backreaction conditions that the excited modes neither spoil the inflationary background nor the slow-roll evolution. The Hadamard condition has automatically been satisfied through our parametrization \eqref{beta}.}
\begin{eqnarray}\label{Backreaction}
|\beta_0^{(s,t)}| \leq \sqrt{\epsilon |\eta'|} \frac{H_\text{inf} \Mpl}{M_{(s,t)}^2}\,,
\end{eqnarray}
with $\eta'$ being the second slow-roll parameter (using the prime so as to avoid confusion with the conformal time parameter $\eta$). From the observed mean value of the spectral tilt of the scalar power spectrum, $1-n_s = 0.0351$ \cite{Planck2}, one can estimate $\eta'$ as 
\begin{eqnarray}\label{Eta}
\eta' \simeq \frac{n_s - 1}{2} + 3\epsilon \approx -0.01755\,, 
\end{eqnarray}
since, for the scalar modes in the BD vacuum, $\epsilon$ has to be a negligibly small number given by \eqref{Eq3}. In writing the above expression, we have assumed that the scale-dependence of the NBD part of the initial state is negligible: more specifically, we have assumed that $\d\log(1+2N_k)/\d \log k < 10^{-2}$, which is valid for our parametrization \eqref{beta} \cite{Ganc:2011dy,Ashoorioon:2013eia}.

Following \cite{Ashoorioon:2018sqb}, from the expression of the curvature power spectrum, we can replace the value of $\Mpl$ in the \eqref{Backreaction} above, specialized for the tensor modes, and rewrite it as
\begin{eqnarray}\label{Tensor_BackReaction}
\left(\frac{M_{(t)}}{H_\text{inf}}\right)^2 \lesssim \frac{\sqrt{|\eta'|}}{\sqrt{8\pi^2 \mathcal{P}_\mathcal{R}}}\, \frac{1}{\beta^{(t)}_0}\,.
\end{eqnarray}
The interesting thing is that the $\epsilon$ from \eqref{Eq2} cancels the one from \eqref{Backreaction}. Plugging in $\mathcal{P}_\mathcal{R} \sim 10^{-9}$ as before, and the value of $\eta'$ from \eqref{Eta}, we get 
\begin{eqnarray}
\left(\frac{M_{(t)}}{H_\text{inf}}\right)^2 \lesssim \frac{500}{\beta_0^{(t)}}\,.
\end{eqnarray}
Since we require that $M_{(t)} > H_\text{inf}$ for the EFT of tensor perturbations to be valid \cite{Holman:2007na}, this gives us an upper limit of the allowed number of excited particles in a given mode $k$, which can be written as
\begin{eqnarray}
\beta^{(t)}_0 < 500\,.
\end{eqnarray}
However, it is easy to see that this allows for a maximum enhancement factor (assuming $\Theta^{(t)} = 0$) of the order $\gamma_t \approx 10^6$ which would still lead to a negligibly small upper bound for $r < 10^{-24}$. This is even after we have been fairly generous in allowing the EFT description to stretch beyond its very extreme limit. Therefore, we conclude that excited tensor modes alone are not sufficient to solve this problem and we also need a NBD state for the scalar perturbations.

Let us now have a NBD state for the scalar perturbations such that $\beta_0^{(s)} \gg 1$. The backreaction condition, analogous to \eqref{Tensor_BackReaction}, for the scalar modes takes the form
\begin{eqnarray}\label{Scalar_BackReaction}
\left(\frac{M_{(s)}}{H_\text{inf}}\right)^2 \lesssim \frac{\sqrt{|\eta'|}}{\sqrt{8\pi^2 \mathcal{P}_\mathcal{R}}}\,  \frac{\sqrt{\gamma_s}}{\beta^{(s)}_0}\,.
\end{eqnarray}
The difference from the tensor case is that since we use the same curvature power spectrum to eliminate $\Mpl$ starting from the relation \eqref{Backreaction} (specialized to the scalar modes), now this relation appears with a $\sqrt{\gamma_{(s)}}$ in the numerator coming from \eqref{NBD_PS_Scal}.  For $\beta_0^{(s)} \gg 1$, $\sqrt{\gamma_{(s)}} \approx \beta_0^{(s)}$ and these factors cancel each other and we get
\begin{eqnarray}
\left(\frac{M_{(s)}}{H_\text{inf}}\right)^2 \lesssim 500\,.
\end{eqnarray}
Therefore, in principle, there is nothing which prevents us from boosting the initial state of the scalar perturbations as much as we want, at least as far as the backreaction constraint is concerned. Of course, there are other constraints on $\beta_0^{(s)}$ coming from the upper bound on the  non-Gaussianity of the curvature bispectrum, but we shall get to that later. For now, let us assume that $|\beta_0^{(s)}|^2 \sim N_0^{(s)} \sim 10^{23}$ for a specific choice of a NBD initial state for scalar perturbations. Together with $\Theta^{(s)} = 0$, \textit{i.e.} allowing for an enhancement in the scalar power spectrum, one gets $\gamma_{(s)} \sim 10^{23}$ and weakens the bound on $\epsilon$ to
\begin{eqnarray}
\epsilon < 10^{-9}\,.
\end{eqnarray}
The bound from \eqref{Eq3}, $\epsilon < 10^{-31}$, is significantly weakened in this case since, for the NBD scalar modes, we have \eqref{Epsilon_NBD} which for a $\gamma_{(s)} \sim 10^{23}$ leads to the above constraint. Of course, as mentioned before, this, by itself, does nothing for our bound on $r$. That is so because, even though $\epsilon$ can now be significantly larger, the expression for $r$ is now further suppressed by the same factor $\gamma_s$ used to enlarge the upper bound on $\epsilon$ and is thus bounded by the same small number.  However, as we shall see, this will allow us to boost the tensor modes a lot more which, in turn, will let us considerably relax the upper bound on $r$.  

But first we must further justify our choice for the very large value of $\beta_0^{(s)}$. If one requires that $M_{(s)} \sim 20 H_\text{inf}$,  the backreaction condition \eqref{Backreaction} for the NBD scalar modes can be written as 
\begin{eqnarray}
\beta_0^{(s)} \lesssim \frac{\Mpl H_\text{inf}}{M_s^2} \sqrt{\epsilon 
	|\eta'|} \lesssim 0.5 \times 10^{18} \times \sqrt{10^{-9 -2}} \lesssim 10^{12}\,.
\end{eqnarray}
This would still allow our choice of very large $N_0^{(s)} \sim 10^{23}$ as shown above. However, the stronger bound on the scalar NBD modes usually comes from the fact that such highly populated excited states typically imply a very large non-Gaussian signal  which would be in conflict with current observations. However, this is where the low value of $\epsilon$ comes to our rescue. The local-type $f_{NL}$, coming from the squeezed configuration $k_1\approx k_2 \gg k_3$, for NBD scalar modes is given by \cite{Holman:2007na,Ganc:2011dy,Agullo:2010ws}
\begin{eqnarray}
f_{NL}^{loc} \sim \frac{5}{3}  \frac{k_1}{k_3} \, \epsilon\,,
\end{eqnarray}
and is independent of the number of particles, $N_0^{(s)}$, in the excited scalar modes when $N_0^{(s)}\gg 1$. This can easily be seen from the explicit expression of the bispectrum for our type of NBD states (with $\Theta^{(s)}=0$) as given in, say, \cite{Ganc:2011dy}. Thus, for $\epsilon < 10^{-9}$, the local-type $f_{NL}$ one gets for these states is completely compatible with Planck data (on assuming $k_1/k_3 \sim 300$). For other configurations  (such as the flattened one with $k_1 \approx k_2 \approx 2 k_3$), the enhancement is much smaller  anyway \cite{Agullo:2010ws}. Thus, we find that there are no obstructions to having a highly populated excited initial state for the scalar perturbations coming from the scalar bispectrum, due to the small values of $\epsilon$ required by the TCC.

Finally, we turn to the last piece of the puzzle to see what are the allowed values of NBD tensor modes, given that $\gamma_{(s)} \neq 1$. For this case, we can rewrite the backreaction condition \eqref{Tensor_BackReaction} for the tensor modes as 
\begin{eqnarray}
\left(\frac{M_{(t)}}{H_\text{inf}}\right)^2 \lesssim \frac{\sqrt{|\eta'|}}{\sqrt{8\pi^2 \mathcal{P}_\mathcal{R}}}\, \frac{\sqrt{\gamma_{(s)}}}{\beta^{(t)}_0}\,.
\end{eqnarray}
Note that the factors of $\sqrt{\gamma_s}$ in the numerator and $\beta^{(t)}_0$ do not cancel for this relation since we assumed different NBD states for the scalar and tensor perturbations. Plugging in $\gamma_{(s)} \sim 10^{23}$, we find that for the EFT description of tensor fluctuations to remain valid, we need
\begin{eqnarray}
\beta_0^{(t)} < 500 \times 10^{11.5}\,.
\end{eqnarray}
Allowing for $M_t \sim 10 H_\text{inf}$ gives us $\gamma_t \sim 10^{28}$. Therefore, plugging in this value in \eqref{r_NBD}, we get an upper bound for the tensor-to-scalar ratio given by
\begin{eqnarray}
r < 0.001\,,
\end{eqnarray}
which is much bigger than the one proposed in \cite{TCC2}. Thus, if one allows for NBD initial states, then the discovery of primordial gravitational waves need not rule out inflation, even assuming the TCC. 

One last check we need to make is regarding the allowed excited states for tensor modes coming from the non-Gaussianity constraints on the tensor bispectrum. It has been shown in the Appendix, following \cite{Ashoorioon:2018sqb}, that for $\gamma_{(t)} \gg 1$, and for $\Theta^{(t)} \sim 0$, the values of $f_{NL}^{hhh}$ both  in the flattened and squeezed configurations are well within current experimental bounds \cite{Shiraishi:2019yux}. Thus, our NBD initial states are completely viable given the current bounds on the tensor bispectrum.

\section{Conclusion}
It has been a relatively new challenge for cosmological models to survive the test of theoretical consistency coming from fundamental quantum gravity principles. Indeed, the so-called dS swampland conjecture was thought to have ruled out many models of inflation at first but, on refinement, it was found that most of the inflationary models could survive it. It is thus somewhat amusing to note that the TCC, which is a weaker condition than the dS conjecture, seems to be able to severely constrain, and perhaps rule out, almost all models of inflation provided we see the discovery of primordial gravitational waves sometime in the future. Indeed, the fact that the TCC is weaker than the dS conjecture has been discussed in detail in \cite{TCC1}, showing that the dS conjecture only follows from it in parametrically large distances in field space. This is precisely the reason why one gets a very small value of $\epsilon$ from the TCC whereas $\epsilon$ was constrained to be $\mathcal{O}(1)$ from the (original) dS conjecture. It is also why the requirement of a small value of $\epsilon$ in \cite{TCC2}, starting from the TCC, is accompanied by small-field excursion of the inflaton when considering a single-field slow-roll model. This brings us to the last bound in \cite{TCC2} put on the field range traversed during inflation, assuming a slow-roll single-field model. This, of course, can perhaps be avoided in multi-field models of inflation or involving other sophisticated dynamics. However, this also shows that the small value of $\epsilon$, at least for a single-field model, is necessarily true only for small field excursions. Therefore, if one can somehow think of a mechanism which allows for a larger field excursion of the inflaton, coupling it to a heavy field or through any other such concoction, this should force one back to larger values for $\epsilon$ since it would be in the regime where one expects to recover (the equivalent of) the dS conjecture.

In this work, we have shown that not only are NBD states more natural to assume for inflation, given the TCC and the requirements of having some pre-inflationary dynamics, they might be one of the few ways of making inflation compatible with both the TCC and the observation of primordial tensor modes. The fundamental mechanism at work here is that we allow the scalar and tensor modes to be in different initial states. All of our calculations were order of magnitude estimates, in the same spirit as \cite{TCC2}, and can easily be changed by small numerical factors either way on more careful examination\footnote{Indeed, the TCC is an approximate statement and it might be that a more rigorous quantum gravity calculation reveals that only modes smaller than $\ell_\text{Pl}/10$ should never cross the Hubble horizon. This would lead to an extra factor of logarithmic corrections. Thus, one should not take the exact value of the upper bound on $r$ very seriously but only the larger message about the order of magnitude.}. Although we require a very large amount of boost for our NBD states so that they can significantly alter the upper bound on $r$, this does not spoil the inflationary background since the scale of inflation has also been significantly reduced from the Planck scale, allowing enough hierarchy for the cut-off scale of the EFT of inflation to be also significantly below $\Mpl$ and yet be above $H_\text{inf}$. The intuitive reason behind this is that the energy density due to the perturbations goes as $\sim \mathcal{O}(H^4)$ whereas the energy density of the background goes as $\sim \mathcal{O}(\Mpl^2 H^2)$. Since the TCC requires that the we can only allow for low-scale inflation, \textit{i.e.} $\Mpl^2 H^2 \gg H^4$, this allows for much larger value of $\beta^{(s,t)}$ than for typical models of inflations at higher energy scales. We also checked that our choice of initial NBD states are compatible with phenomenological constraints.  Although we needed to significantly fine-tune our initial state so as to reach the constraint $r < 10^{-3}$, nevertheless this gives a proof of principle that there exists part of the parameter space of inflation which can allow for detection of primordial tensor modes and yet be compatible with the TCC.  

Looking ahead, this analyses can be put on firmer ground if a \textit{physical} mechanism can be invoked which can generate these kind of initial states which were required for our solution. We hope to pursue this in future work.

\section*{Acknowledgments}
I am thankful to Robert Brandenberger for extensive discussions and introductory lectures on the TCC, and for many useful suggestions on an earlier version of this draft. This research is supported in part by funds from NSERC, from the Canada Research Chair program and by a McGill Space Institute fellowship.

\section*{Appendix: Tensor bispectrum for NBD states}
Here, we quote a few results from \cite{Ashoorioon:2018sqb} regarding the amplitude of the $f_{NL}^{hhh}$ one gets for excited tensor modes, adapted to our conventions. The expression for the flattened configuration $(k_1 +k_3 \approx k_2)$ is given by (keeping only the leading order term)
\begin{eqnarray}
f_{NL}^{hhh,\,\text{flat}}\sim \left(\frac{k_1 k_2 k_3 (k_1^2+k_1 k_3+k_3^2)}{8 \left(k_1^3+(k_1+k_3)^3+k_3^3\right)}  \left(-\frac{1}{2} (N_0^{(t)})^2 + \frac{1}{2} \right) \eta_0^2 +\mathcal{O}(\eta_0) + \mathcal{O}(\eta_i^0) \right)\,\left(N_0^{(t)}\right)^{-2}\,.
\end{eqnarray}
Firstly, we see that  for $\gamma_{(t)} \gg 1$, the $f_{NL}^{hhh,\,\text{flat}}$ is independent of the number of excited particles,where we have assumed $\Theta^{(t)} \sim 0$ like before, just as in the scalar case. In this case, on using the relation $\eta_0 = - M_{(t)}/(k_1 H_\text{inf})$, and assuming $M_{(t)} \sim 10 H_\text{inf}$ as for our model, we find that $f_{NL}^{hhh,\,\text{flat}} \sim \mathcal{O}(10)$ even on assuming a shape such as $k_1 =0.25 k_3 = 0.2 k_2$, which enhances the momenta-dependent factor, as discussed in \cite{Ashoorioon:2018sqb}. In this case, it is easy to verify that for $M_t/H \sim 10$, the maximum value of $f_{NL}^{hhh,\,\text{flat}}$ is small and well within the current bounds \cite{Shiraishi:2019yux}.

For the local configuration, $k_3\ll k_1\approx k_2$, the expression for $f_{NL}^{hhh,\,\text{loc}}$ takes the form \cite{Ashoorioon:2018sqb}
\begin{eqnarray}
f_{NL}^{hhh,\,\text{loc}} \sim -\frac{k_1}{16 k_3} \,\left(N_0^{(t)}\right)^{-2}\,,
\end{eqnarray}
which has, once again, simplified considerably for our case of $\Theta^{(t)} \sim 0$. This shows that $f_{NL}^{hhh,\,\text{loc}}$ is suppressed by factors of $N_0^{(t)}$ and is negligibly small for our highly excited state.

\end{document}